\pdfoutput=1
\documentclass[prl,twocolumn,showpacs,amsmath,amssymb,floatfix,superscriptaddress]{revtex4}

\usepackage{graphicx}
\usepackage{epsfig}
\usepackage{dcolumn}
\usepackage{bm}

\begin{document}

\title{Spin-transfer torque and spin-polarization in topological-insulator$|$ferromagnet vertical heterostructures}

\author{Farzad Mahfouzi}
\affiliation{Department of Physics and Astronomy, University of Delaware, Newark, DE 19716-2570, USA}
\author{Naoto Nagaosa}
\affiliation{Cross-Correlated Materials Research Group (CMRG) and Correlated Electron Research Group (CERG), RIKEN-ASI, Wako, Saitama 351-0198, Japan}
\affiliation{Department of Applied Physics, University of Tokyo, Tokyo 113-8656, Japan}
\author{Branislav K. Nikoli\' c}
\email{bnikolic@udel.edu}
\affiliation{Department of Physics and Astronomy, University
of Delaware, Newark, DE 19716-2570, USA}
\affiliation{Cross-Correlated Materials Research Group (CMRG) and Correlated Electron Research Group (CERG), RIKEN-ASI, Wako, Saitama 351-0198, Japan}

\begin{abstract}
We predict an {\em unconventional} spin-transfer torque (STT) acting on the magnetization of a free ferromagnetic (F) layer within N$|$TI$|$F vertical heterostructures which originates from strong spin-orbit coupling (SOC) on the surface of a three-dimensional topological insulator (TI), as well as from charge current becoming spin-polarized in the direction of transport as it flows from the normal metal (N) across the bulk of the TI slab. Unlike conventional STT in symmetric F$^\prime|$I$|$F magnetic tunnel junctions, where only the in-plane STT component is non-zero in the linear response, both the in-plane and perpendicular torque are sizable in N$|$TI$|$F junctions while {\em not requiring} fixed F$^\prime$ layer as spin-polarizer which is advantageous for spintronic applications. Using the nonequilibrium Born-Oppenheimer treatment of interaction between fast conduction electrons and slow magnetization, we derive a general Keldysh Green function-based STT formula which makes it possible to analyze torque in the presence of SOC either in the bulk or at the interface of the free F layer.
\end{abstract}

\maketitle
 
The spin-transfer torque (STT) is a phenomenon in which spin current of large enough density injected into a ferromagnetic (F) layer either switches its magnetization from one static configuration to another or generates a dynamical situation with steady-state precessing magnetization~\cite{Ralph2008,Gambardella2011}. The origin of STT is absorption of itinerant flow of angular momentum components normal to the magnetization direction. It represents one of the {\em central phenomena} of the second-generation spintronics, focused
on manipulation of coherent spin states, since reduction of current densities (currently of the order 10$^6$-10$^8$ A/cm$^2$)  required for STT-based magnetization switching is expected to bring commercially viable magnetic random access memories~\cite{Katine2008}. The rich nonequilibrium physics~\cite{Wang2011} arising in the interplay of spin currents carried by fast conduction electrons and collective magnetization dynamics, viewed as the slow classical degree of freedom, is of great fundamental interest.

The early phenomenological explanations~\cite{Slonczewski1996} of STT in noncollinear ferromagnetic metal circuits have been followed by more microscopic theories~\cite{Stiles2002,Edwards2005,Haney2007,Wang2008b,Tang2010,Xiao2008a}, which are often combined with first-principles input about real materials~\cite{Stiles2002,Edwards2005,Haney2007,Wang2008b}. These theories have been focused on devices with no spin-orbit coupling (SOC) where STT is directly connected to the divergence of spin current as a consequence of the conservation of total spin. Thus, STT vector can be obtained simply from the local spin current at the N$|$F or I$|$F interface (N-normal metal, I-insulating barrier) within F$^\prime|$N$|$F spin valves or F$^\prime|$I$|$F  magnetic tunnel junctions (MTJs). Such local spin currents are typically computed using the Landauer-B\"{u}tikker scattering approach~\cite{Stiles2002,Wang2008b} or the nonequilibrium Green function (NEGF) formalism~\cite{Edwards2005,Haney2007,Tang2010}.

\begin{figure}
\includegraphics[scale=0.4,angle=0]{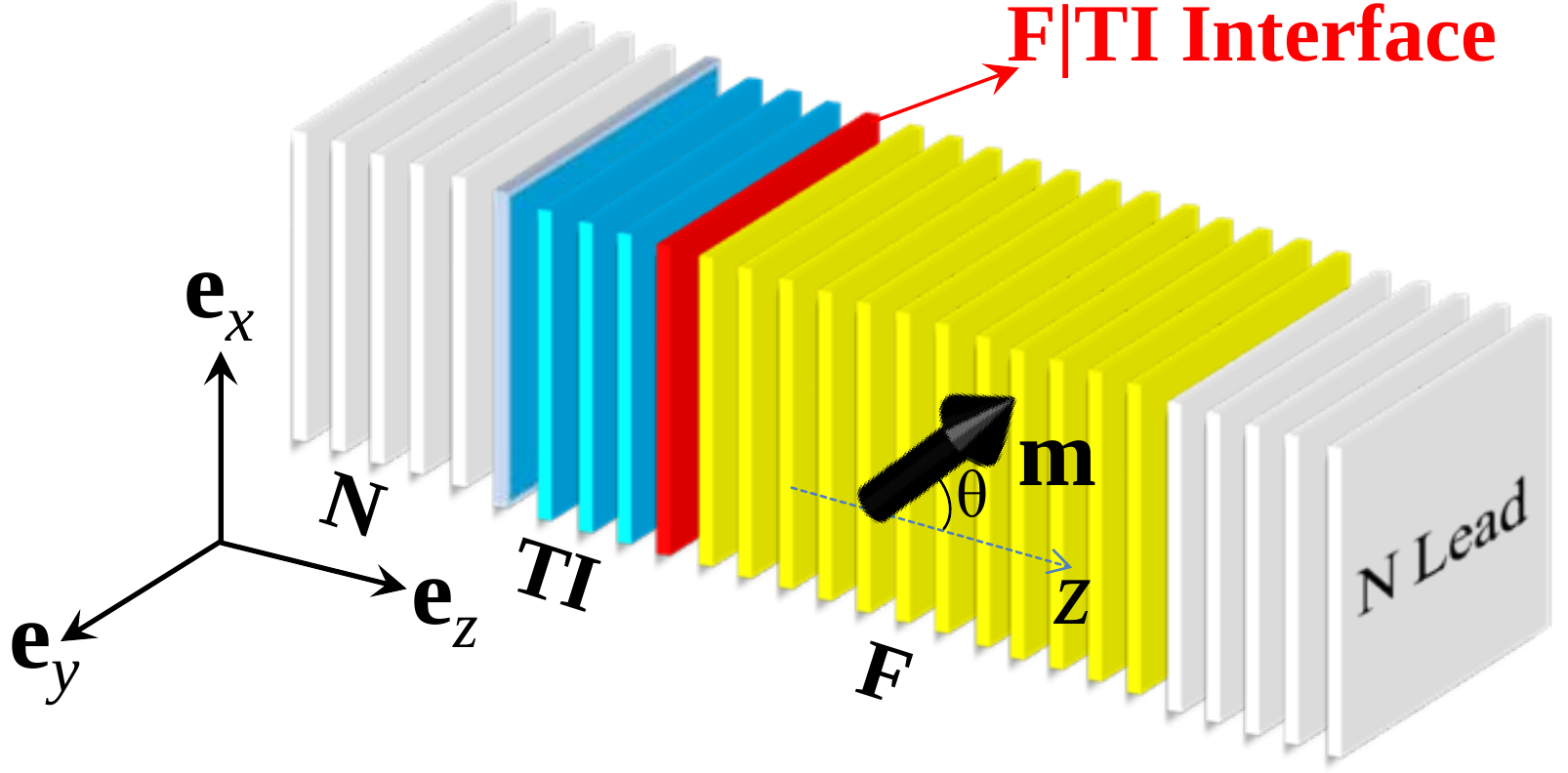}
\caption{(Color online) Schematic view of the topological-insulator-based vertical heterostructure operated by spin-transfer torque. The junction  contains a {\em single} free F layer of finite thickness with magnetization ${\bf m}$, which is attached to the right semi-infinite N lead. Unpolarized charge current is injected from the left semi-infinite N lead across finite thickness TI slab. We assume that each layer is composed of atomic monolayers modeled on an infinite square tight-binding lattice.}
\label{fig:setup}
\end{figure}

However, this methodology is {\em inapplicable} to junctions with {\em strong} SOC, which has recently ignited intense theoretical efforts~\cite{Haney2007,Manchon2008,Hals2010,Haney2010,Manchon2011} to devise approaches for efficient computation of STT in the presence of spin non-conserving interactions. For example, SOC can be introduced into the device by the bulk ferromagnets (as in F layers based on ferromagnetic semiconductors~\cite{Gambardella2011,Hals2010,Haney2010}), or due to the Rashba SOC at the F$|$I interface in devices with structural inversion asymmetry~\cite{Gambardella2011}. The importance of the latter for potential applications was demonstrated in the very recent experiment~\cite{Miron2010} measuring ``SO torque''~\cite{Gambardella2011,Manchon2008} in Pt$|$Co$|$AlO$_x$ semi-MTJ where charge current flows within the plane of a Co layer.

Concurrently, the recent discovery~\cite{Hasan2010} of three-dimensional (3D) topological insulators (TIs), which possess a usual band gap in the bulk while hosting metallic surfaces whose massless Dirac electrons have spins locked with their momenta due to the strong Rashba-type SOC, has led to theoretical proposals to employ such exotic states of matter for STT applications. For example, magnetization of a  ferromagnetic film with perpendicular anisotropy deposited on TI surface could be switched by interfacial quantum Hall current~\cite{Garate2010}.

However, very little is known about the device geometries~\cite{Mahfouzi2011} in which charge and spin currents are {\em perpendicular} to the surface of TI and their potential for applications in conventional~\cite{Wang2011} vertical MTJ setups. In general, vertical TI-based heterostructures would exploit strong interfacial SO coupling {\em without requiring}~\cite{Mahfouzi2011,Zhao2010} perfectly insulating bulk whose unintentional doping in present experiments obscures~\cite{Butch2010} topological properties anticipated for lateral transport along the TI surface.

\begin{figure}
\includegraphics[scale=0.34,angle=0]{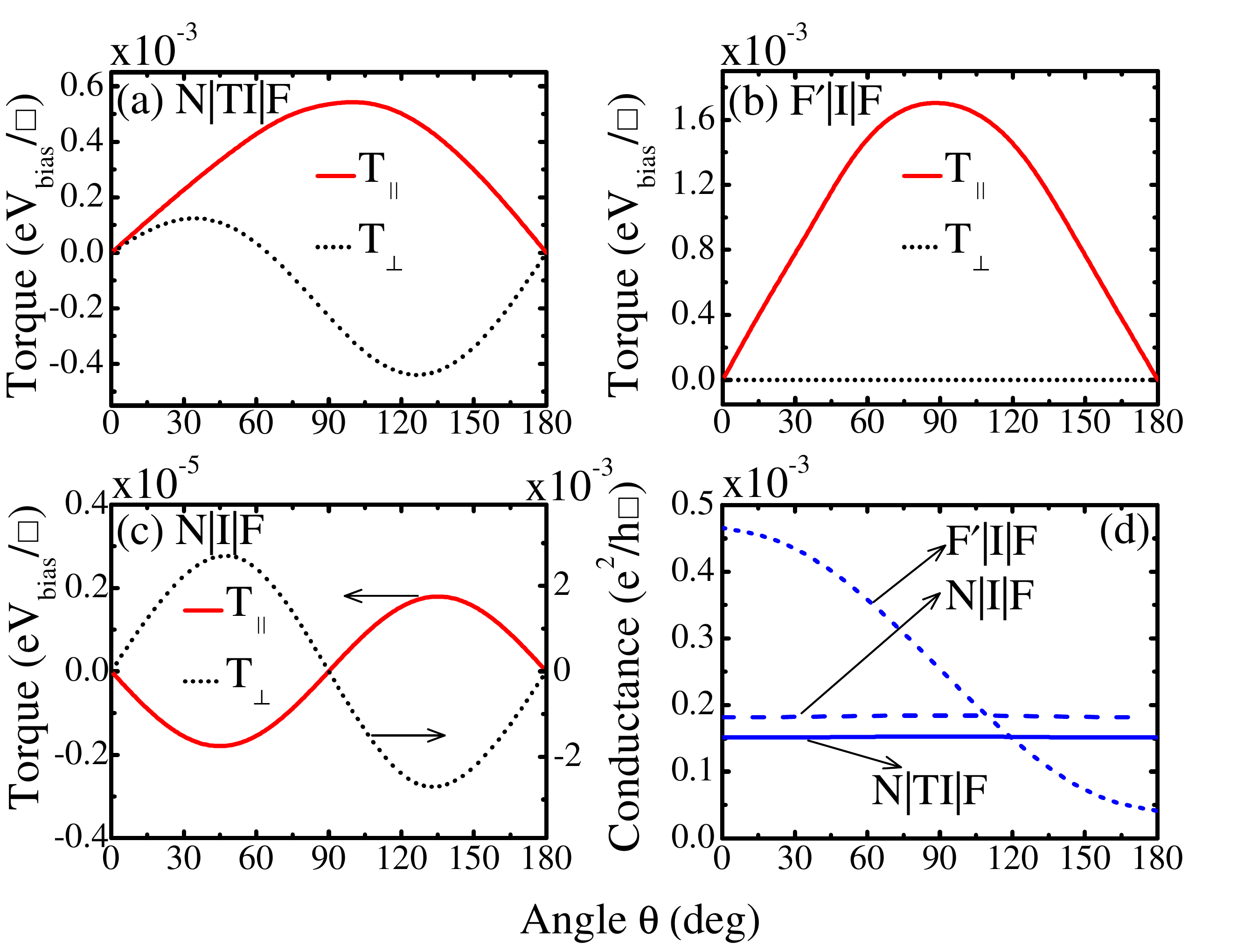}
\caption{(Color online) (a) The angular dependence of STT components, ${\bf T}_{\parallel} = \tau_{\parallel} {\bf m} \times ({\bf m} \times {\bf e}_z)$ and ${\bf T}_{\perp}=\tau_{\perp} {\bf m} \times {\bf e}_z$, acting on the free-layer magnetization ${\bf m}$ in N$|$TI$|$F semi-MTJ shown in Fig.~\ref{fig:setup}. (b) The STT components, ${\bf T}_{\parallel} = \tau_{\parallel} {\bf m} \times ({\bf m} \times {\bf m}^\prime)$ and ${\bf T}_{\perp}=\tau_{\perp} {\bf m} \times {\bf m}^\prime$, acting on the free-layer magnetization \mbox{${\bf m}=(\sin \theta \cos \phi, \sin \theta \sin \phi, \cos \theta)$} in conventional F$^\prime|$I$|$F symmetric MTJ where magnetization of the reference layer F$^\prime$  is fixed at ${\bf m}^\prime={\bf e}_z$. (c) The STT components in N$|$I$|$F semi-MTJ, defined in the same fashion as in panel (a), with the Rashba SOC of strength \mbox{$\alpha_R/2a=0.1$ eV} located on the monolayer of F which is in contact with I barrier. (d) The angular dependence of conductances for N$|$TI$|$F, F$^\prime|$I$|$F and N$|$I$|$F junctions. The bias voltage $V_b$ in all panels is sufficiently small to ensure the linear-response regime.}
\label{fig:fig2}
\end{figure}

In this Letter, we  derive an efficient NEGF (in Keldysh formulation)-based formula which makes it possible to analyze STT in the presence of arbitrary SOC within the device. Unlike the recent formulas~\cite{Hals2010,Haney2010} developed to treat SOC effects on STT in the linear-response regime, ours can handle torque driven by finite bias voltage (required to reach sufficient current density in MTJs~\cite{Wang2011}), and  it can also be easily combined with density functional theory (DFT) through the NEGF-DFT formalism~\cite{Haney2007,Jia2011}. This STT formula is then applied to predict unusual features, shown  in Fig.~\ref{fig:fig2}(a), of the in-plane ${\bf T}_{\parallel}$ torque emerging in TI-based semi-MTJ illustrated in Fig.~\ref{fig:setup} in the {\em absence} of any external spin-polarizer. For conventional F$^\prime|$I$|$F MTJs, where F$^\prime$ reference layer with magnetization ${\bf m}^\prime$ plays the role of an external spin-polarizer, it is customary to analyze the in-plane (originally considered by Slonczewski~\cite{Slonczewski1996})  and perpendicular  (or out-of-plane)  torque components, \mbox{${\bf T} = {\bf T}_{\parallel} + {\bf T}_{\perp}$}. The in-plane component \mbox{$\mathbf{T}_{\parallel}=\tau_{\parallel}{\bf m} \times ({\bf m} \times {\bf m}^\prime)$}  is purely nonequilibrium and competes with the damping.  The perpendicular torque \mbox{$\mathbf{T}_{\perp}=\tau_{\perp} {\bf m} \times {\bf m}^\prime$}  arises from spin reorientation at the interfaces and possesses both equilibrium (i.e., interlayer exchange coupling) and nonequilibrium components which act like an effective magnetic field on the magnetization ${\bf m}$ of the free F layer~\cite{Xiao2008a}. While ${\bf T}_{\perp}$ component is vanishingly small in metallic spin valves~\cite{Edwards2005,Wang2008b}, it can be substantial~\cite{Wang2011} in MTJs due to the momentum filtering imposed by the tunnel barrier~\cite{Tang2010,Xiao2008a}.

\begin{figure}
\includegraphics[scale=0.25,angle=0]{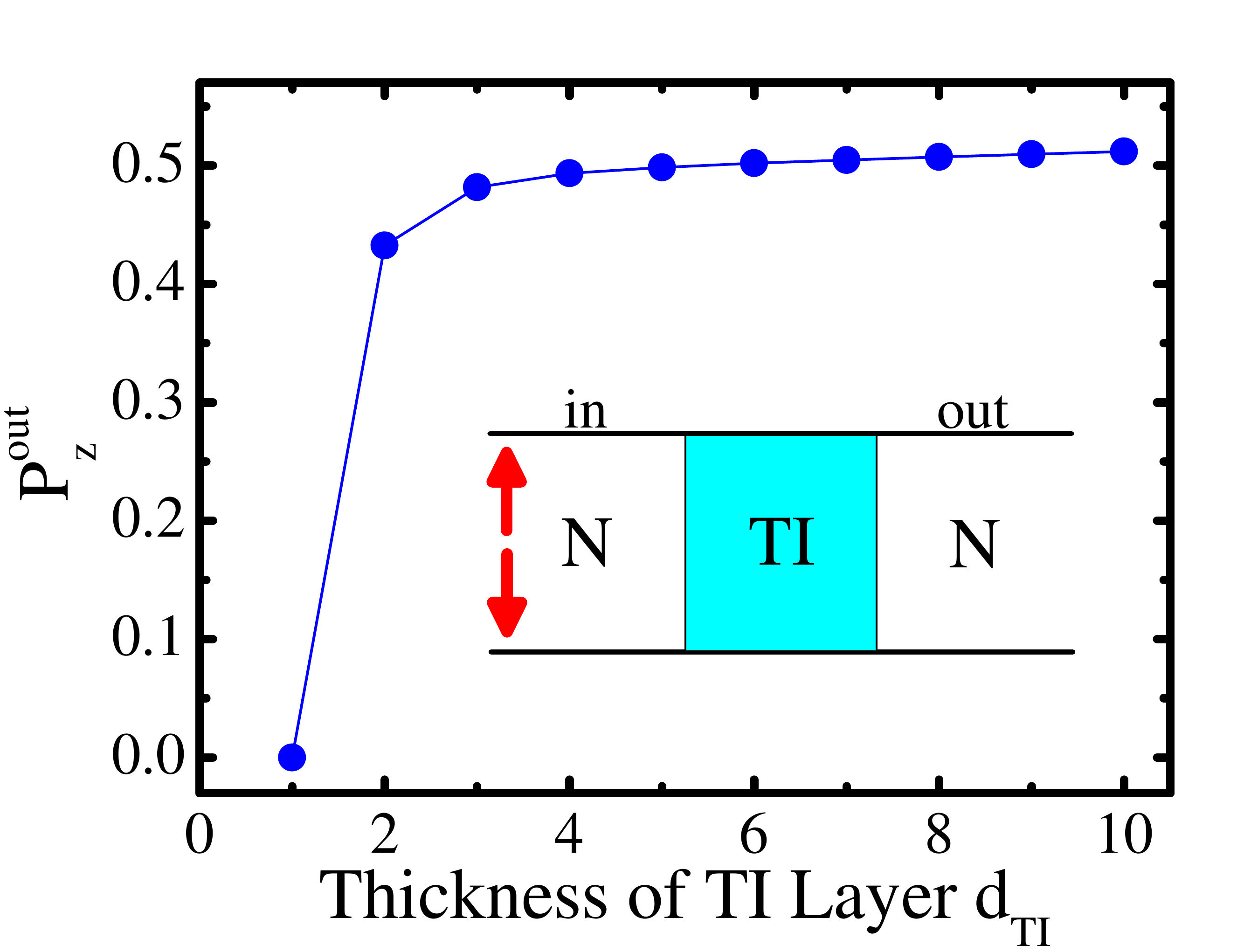}
\caption{(Color online) The spin-polarization vector \mbox{${\bf P}^{\rm out}=(0,0,P^{\rm out}_z)$} of current~\cite{Nikolic2005} in the right N lead of N$|$TI$|$N junction as a function of the thickness $d_{\rm TI}$ of the 3D TI layer after unpolarized charge current is injected from the left N lead.}
\label{fig:fig3}
\end{figure}

To elucidate the effect of TI slab on unpolarized charge current injected from the left N lead we analyze the spin density matrix \mbox{$\hat{\rho}^{\rm out}_{\rm spin} = \frac{1}{2}(1+{\bf P}^{\rm out}\cdot \hat{\bm \sigma})$} for an ensemble of outgoing spin in the right N lead of N$|$TI$|$N junction. The expression for $\hat{\rho}^{\rm out}_{\rm spin}$, or equivalent spin-polarization vector ${\bf P}^{\rm out}$, was derived as Eq.~(10)  in Ref.~\cite{Nikolic2005} in terms of the scattering matrix of the device. Its evaluation for N$|$TI$|$N junction is plotted in Fig.~\ref{fig:fig3}, which shows how TI slab  polarizes the incoming current in the direction of transport with \mbox{${\bf P}^{\rm out}=(0,0,\simeq 0.5)$}. The polarizing effect of the TI slab comes from the effective momentum-dependent magnetic field along the $z$-axis [encoded by the ${\bm \Gamma}_3$ term in the TI Hamiltonian in Eq.~(\ref{eq:ti}) discussed below], which requires sufficient thickness of the slab to manifest itself as well as that the Fermi energy of the device $E_F$ is within the bulk gap of the TI slab. The spin-polarization of charge current induced by its flow through a finite-size region with SOC has been discussed previously for low-dimensional systems (such as the two-dimensional electron gas with the Rashba SOC~\cite{Adagideli2006}). Due to the constraints imposed by the time-reversal invariance, such SOC-induced polarization cannot~\cite{Adagideli2006} be detected via current or voltage measurement on standard two-terminal ferromagnetic circuits, as exemplified by  Fig.~\ref{fig:fig2}(d) where conductance of N$|$TI$|$F junction is the same for ${\bf m} \parallel {\bf e}_z$ and  ${\bf m} \nparallel {\bf e}_z$ configurations.

Following this analysis, the meaning of torque components plotted in Fig.~\ref{fig:fig2}(a) for semi-MTJ is defined by
\begin{equation}\label{eq:sttcomponents}
{\bf T} = \mathbf{T}_{\parallel}+\mathbf{T}_{\perp} = \tau_{\parallel} {\bf m} \times ({\bf m} \times {\bf e}_z) + \tau_{\perp} {\bf m} \times {\bf e}_z.
\end{equation}
In fact, the same definition of torque components is applicable~\cite{Manchon2011}  to N$|$I$|$F semi-MTJ with the strong Rashba SOC, $\alpha_R (\hat{\bm \sigma} \times {\bf k}_{\parallel})\cdot {\bf e}_z$, at the I$|$F interface~\cite{Gambardella2011,Miron2010}. In that case, ${\bf T}_\parallel$ and ${\bf T}_\perp$ components plotted in Fig.~\ref{fig:fig2}(c) are driven purely by the surface Rashba SOC, which is the second order effect $\propto \alpha_R^2$ characterized by torque asymmetry~\cite{Manchon2011} around the stable magnetic state $\theta=90^\circ$. In contrast, ${\bf T}_\parallel$  in N$|$TI$|$F semi-MTJ are non-zero at $\theta = 90^\circ$ due to the summation of asymmetric contribution driven by the strong SOC on the surface of TI layer and symmetric [akin to torque in the usual MTJ shown in Fig.~\ref{fig:fig2}(b)] contribution generated by the conventional STT due to spin-polarization of the current passing through the bulk of the TI layer. 
The existence of $\mathbf{T}_{\parallel}$ and $\mathbf{T}_{\perp}$ for N$|$I$|$F or N$|$TI$|$F semi-MTJs makes this type of STT quite different from SO torques~\cite{Gambardella2011,Manchon2008,Haney2010} that act only as an effective magnetic field which can induce switching {\em but not precession} of the magnetization in the free F layer~\cite{Manchon2011}.

The exploitation~\cite{Katine2008} of STT in conventional F$^\prime|$I$|$F MTJs demands a compromise between large current density (requiring low junction resistance to avoid damage) and readability (requiring large magnetoresistance). In addition, optimization of spin polarization across the junction, stabilization of the fixed magnetization of the reference F$^\prime$ layer and minimization of stray fields demands complex stacking structure (involving typically more than ten different layers~\cite{Ralph2008,Gambardella2011,Katine2008,Wang2011}). On the other hand, our semi-MTJ requires only one F layer. Furthermore, Figs.~\ref{fig:fig2}(a),(b) show that $\mathbf{T}_{\parallel}$ in N$|$TI$|$F semi-MTJ is comparable to the one  in F$^\prime|$I$|$F MTJ tuned (via the on-site potential in the I layer) to have similar resistance. The angular dependence of conductances for  N$|$TI$|$F, N$|$I$|$F, and F$^\prime|$I$|$F junctions are compared in Fig.~\ref{fig:fig2}(d).

Now we turn to the detailed explanation of our formalism. The semi-MTJ in Fig.~\ref{fig:setup} is modeled on a cubic lattice, with lattice constant $a$ and unit area $a^2 \equiv \Box$, where monolayers of different materials (N, F, TI)  are infinite in the transverse $xy$-direction. The TI layer has thickness $d_{\rm TI}=5$ and the free F layer has thickness $d_F=70$ monolayers. The F and N layer are described by a tight-binding Hamiltonian with a single $s$-orbital per site
\begin{eqnarray} \label{eq:fnhamiltonian}
\hat{H}_F & = & \sum_{n,\sigma\sigma',{\bf k}_{\parallel}} \hat{c}_{n\sigma,{\bf k}_{\parallel}}^{\dagger}\left(\varepsilon_{n,{\bf k}_{\parallel}} \delta_{\sigma\sigma'} - \frac{\Delta_n}{2} \mathbf{m} \cdot [\hat{\bm \sigma}]_{\sigma \sigma'} \right) \hat{c}_{n\sigma',{\bf k}_{\parallel}} \nonumber \\
&& \mbox{} - \gamma \sum_{n,\sigma,{\bf k}_{\parallel}}  (\hat{c}_{n\sigma,{\bf k}_{\parallel}}^{\dagger}\hat{c}_{n+1, \sigma,{\bf k}_{\parallel}} + \mathrm{H.c.}).
\end{eqnarray}
The operators $\hat{c}_{{\bf n}\sigma}^\dag$ ($\hat{c}_{{\bf n}\sigma}$) create (annihilate) electron with spin $\sigma$ on monolayer $n$ with transverse momentum ${\bf k}_{\parallel}$ within the monolayer. The in-monolayer kinetic energy is $\varepsilon_{n,{\bf k}_{\parallel}} = - 2\gamma(\cos k_y a + \cos k_z a)$, whose effect is equivalent to an increase in the on-site energy, and the nearest neighbor hopping is \mbox{$\gamma=1.0$ eV}. The coupling of itinerant electrons to collective magnetization dynamics is described through the material-dependent exchange potential \mbox{$\Delta_n=1.0$ eV} ($\Delta_n \equiv 0$ within semi-infinite ideal N leads), where $\hat{\bm \sigma}=(\hat{\sigma}_x,\hat{\sigma}_y,\hat{\sigma}_z)$ is the vector of the Pauli matrices and $[\hat{\sigma}_\alpha]_{\sigma\sigma'}$ denotes the Pauli matrix elements.

The minimal model for the slab of 3D TI, such as Bi$_2$Se$_3$, is the effective tight-binding Hamiltonian with four orbitals per site~\cite{Liu2010}:
\begin{eqnarray}\label{eq:ti}
\begin{split}
& \hat{H}_{\rm TI}  =  \sum_{n, {\bf k}_{\parallel}} \left\{ {\bf c}^\dagger_{n,{\bf k}_{\parallel}} \left(\frac{B}{a^2} {\bm \Gamma}_0 - i\frac{A}{2a} {\bm \Gamma}_3 \right) {\bf c}_{n+1,{\bf k}_{\parallel}} + \mathrm{H.c.} \right. \nonumber \\
&+  \left. {\bf c}^\dagger_{n,{\bf k}_{\parallel}} \left[ C\bm{1} + d({\bf k}_{\parallel}) {\bm \Gamma}_0 + \frac{A}{a}({\bm \Gamma}_1 \sin k_x a + {\bm \Gamma}_2 \sin k_y a) \right] {\bf c}_{n,{\bf k}_{\parallel}} \right\}.
\end{split}
\end{eqnarray}
It yields the correct gap size in the bulk and surface dispersion while reducing to the continuum ${\bf k} \cdot {\bf p}$ Hamiltonian in the small $k$ limit. Here $\hat{\bf c}=(\hat{c}_{+\uparrow}, \hat{c}_{+\downarrow}, \hat{c}_{-\uparrow}, \hat{c}_{-\downarrow})^T$ annihilates electron in different orbitals, $d({\bf k})_{\parallel} = M - 2B/a^2 + 2B(\cos k_xa + \cos k_ya -2)/a^2$, ${\bm \Gamma}_i$ $(i=0,1,2,3)$ are $4\times 4$ Dirac matrices and $\bm{1}$ is the unit matrix of the same size. The numerical values of parameters are chosen as: \mbox{$M=0.3$ eV}; \mbox{$A=0.5$ $a$eV}; and \mbox{$B=0.25$ $a^2$eV}. The Fermi energy of the whole device is set at \mbox{$E_F=3.1$ eV}, and the bottom of the band of the TI layer is shifted by \mbox{$C=3.0$ eV}.

The hopping \mbox{$\gamma_c=0.25$ eV} between the sites of metallic F or N layer and the TI layer is chosen to ensure that the Dirac cone on the surface of TI is not distorted~\cite{Mahfouzi2011,Zhao2010} by the penetration of evanescent modes from these neighboring metallic layers. The weak F to TI coupling can be achieved by growing an ultrathin layer of a conventional band insulator, such as  In$_2$Se$_3$ with large bandgap and good chemical and structural compatibility with Bi$_2$Se$_3$ where sharp heterointerfaces have already been demonstrated by molecular-beam epitaxy growth~\cite{Wang2011a}. We assume that such layer of sufficient thickness is present and it suppresses the magnetic proximity effect, i.e., $\Delta_n=0$ on the TI monolayer (denoted as F$|$TI interface in Fig.~\ref{fig:setup}) closest to the F layer.

Using the operators $\hat{c}_{n \sigma}^\dag$ ($\hat{c}_{n \sigma}$) which create (annihilate) electron with spin $\sigma$ on monolayer $n$, we can introduce the two fundamental objects~\cite{Haug2007} of the NEGF formalism---the retarded  $G^{r,\sigma\sigma'}_{nn'}(t,t')=-i \Theta(t-t') \langle \{\hat{c}_{n\sigma}(t) , \hat{c}^\dagger_{n'\sigma'}(t')\}\rangle$ and the lesser $G^{<,\sigma\sigma'}_{nn'}(t,t')=i \langle \hat{c}^\dagger_{n'\sigma'}(t') \hat{c}_{n \sigma}(t)\rangle$ GF that describe the density of available quantum states and how electrons occupy those states, respectively. Here $\langle \ldots \rangle$ denotes the nonequilibrium statistical average~\cite{Haug2007}. In stationary problems, $\hat{G}^r$ and $\hat{G}^<$ depend only on the time difference $t-t^\prime$ or energy $E$ after the Fourier transform.

In the absence of SOC, one can obtain STT in F$^\prime|$N$|$F spin valves or F$^\prime|$I$|$F MTJs by computing~\cite{Tang2010} the vector of spin current between two neighboring monolayers $n$ and $n+1$  coupled by the hopping parameter $\gamma$:
\begin{equation}\label{eq:localcurrent}
{\bf I}^S_{n,n+1} = \frac{\gamma}{4\pi} \int dE d\mathbf{k}_{\parallel} \, \mathrm{Tr}_\sigma \, [\bm{\sigma} (G^{<,\sigma\sigma'}_{n+1,n} -  G^{<,\sigma\sigma'}_{n,n+1})].
\end{equation}
The integration over $\mathbf{k}_{\parallel}$ is required because of the device translational invariance in the transverse direction. Since for conserved spin current, the monolayer-resolved~\cite{Wang2008b} STT  is given by ${\bf T}_{n} = -\nabla \cdot {\bf I}^S = {\bf I}_{n-1,n}^S -  {\bf I}_{n,n+1}^S$, the total torque on the free F layer is~\cite{Tang2010},  ${\bf T} = \sum_{\lambda'=0}^\infty  ({\bf I}_{n-1,n}^S - {\bf I}_{n,n+1}^S) = {\bf I}_{-1,0}^S - {\bf I}_{\infty,\infty}^S = \mathbf{I}^S_{-1,0}$. Here the subscripts -1 and 0 refer to the last monolayer of the N or I barrier and the first monolayer of the F layer, respectively. In the multilayers with SOC such as those in Fig.~\ref{fig:setup}, this straightforward NEGF strategy to get STT {\em becomes inapplicable} since spin current will not decay (i.e., ${\bf I}^S_{\infty,\infty} \neq 0$) if SOC is present in the bulk of the free F layer~\cite{Haney2010}. Also,  spin current across the interface  $\mathbf{I}^S_{-1,0}$ is {\em insufficient} to get STT if strong SOC is present directly at the interface.

To derive a general NEGF-based expression for a mean current-induced force, we start by assuming that the device Hamiltonian depends on a variable
$q$ which corresponds to  slow collective (i.e., ``mechanical'') degrees of freedom. The expectation value of the corresponding canonical force \mbox{$\hat{Q} = -\partial \hat{H}/\partial q$} is obtained using the density matrix $\hat{\rho} = \int dE\, \hat{G}^{<}(E,q)$:
\begin{eqnarray}\label{eq:nebo}
Q =  -\frac{1}{2\pi i} \int\limits_{-\infty}^{+\infty} dE\, \mathrm{Tr}\, \left[\frac{\partial \hat{H}}{\partial q}\hat{G}^{<} \right] =  -\left \langle\frac{\partial \hat{H}}{\partial q}\hat{G}^{<} \right \rangle,
\end{eqnarray}
where $\hat{G}^{<}(E,q)$ is adiabatic GF obtained for a frozen-in-time variable $q$.  By exchanging the derivative between the Hamiltonian and $\hat{G}^{<}(E,q)$, $Q =-\partial \langle\hat{H}\hat{G}^{<}\rangle /\partial q +\langle\hat{H} \partial \hat{G}^{<}/ \partial q \rangle$, and by using the standard equations for the retarded and lesser GFs~\cite{Haug2007}, $\hat{G}^{r}(E) = [E\hat{I} - \hat{H} - \hat{\Sigma}^r]^{-1}$ and $\hat{G}^{<}(E)  =   \hat{G}^{r}(E) \hat{\Sigma}^{<}(E) \hat{G}^{a}(E)$, we finally obtain
\begin{equation} \label{eq:central}
Q = i \left \langle\frac{\partial \hat{G}^r}{\partial q}\hat{\Sigma}^{<}\hat{G}^a\hat{\Gamma}\right \rangle - \left \langle \hat{\Sigma}^{<} \frac{\partial \hat{G}^r}{\partial q} \right \rangle.
\end{equation}
The advanced GF is given by $\hat{G}^a=[\hat{G}^r]^\dagger$ and $\hat{I}$ is the unit operator. For devices where electron-electron or electron-phonon interactions can be neglected, $\hat{\Sigma}^r(E)=\sum_p \hat{\Sigma}^r_p(E-eV_p)$ is the sum of retarded self-energies due to the coupling to semi-infinite ideal (F or N) leads  $p=L,R$, $\hat{\Gamma}_p(E-eV_p)=i [\hat{\Sigma}^{r}_p(E-eV_p)-\hat{\Sigma}^{a}_p(E-eV_p)]$ is the level broadening operator, and \mbox{$\hat{\Sigma}^<(E) = \sum_p i f_p(E) \hat{\Gamma}_p(E-eV_p)$} is the  lesser self-energy~\cite{Haug2007}. The junction is biased by the voltage $eV_b = eV_L - eV_R$ and $f_p(E)=f(E-eV_p)$ is the Fermi function of the macroscopic reservoir to which the lead $p$ is assumed to be attached at infinity.

The expression Eq.~(\ref{eq:central}) is the central formula of our formalism. We note that this STT formula is akin to the mean value of time-averaged force  in nonequilibrium Born-Oppenheimer approach~\cite{Lu2009} to current-induced forces exerted by conduction electrons on ions in nanojunctions or mechanical degrees of freedom in nanoelectromechanical systems whose collective modes are slow compared to electronic time scales. The application of Eq.~(\ref{eq:central}) to get $T_\alpha$ ($\alpha=x,y,z$) component of the STT vector acting on the magnetization of the free F layer within N$|$TI$|$F junction proceeds by first computing $\hat{G}^r(E)$ for the device described by the Hamiltonian $\hat{H} = \hat{H}_{\rm TI} + \hat{H}_{F}$. In the second step, the Hamiltonian of the F layer is modified
\begin{equation}\label{eq:hfprime}
\hat{H}_{F}^q = \hat{H}_{F} + q\sum_{n,\sigma\sigma',{\bf k}_{\parallel}}  \hat{c}_{n\sigma,{\bf k}_{\parallel}}^{\dagger} [{\bf e}_\alpha  \cdot ({\bf m} \times \hat{\bm \sigma})]_{\sigma\sigma'} \hat{c}_{n\sigma',{\bf k}_{\parallel}},
\end{equation}
and $\hat{G}^r(E)[\hat{H}^q]$ is computed for the new Hamiltonian $\hat{H}^q  = \hat{H}_{\rm TI} + \hat{H}_{F}^q$. This allows us to obtain $\partial \hat{G}^r/\partial q \approx (\hat{G}^r[\hat{H}^q]-\hat{G}^r[\hat{H}])/q$ where we use $q = 10^{-7}$ as the infinitesimal. The derivative $\partial \hat{G}^r/\partial q$ plugged into Eq.~(\ref{eq:central}) yields $Q = T_\alpha$.


Equation~(\ref{eq:central}) includes both the equilibrium \mbox{${\bf T}_\perp(V_b=0)$}~\cite{Haney2007,Xiao2008a,Tang2010} and experimentally measured~\cite{Wang2011}  {\em nonequilibrium} \mbox{${\bf T}_\perp(V_b)-{\bf T}_\perp(V_b=0)$} contribution to ${\bf T}_\perp$. The linear-response contribution can be extracted by expanding the density matrix $\hat{\rho}$ to first order in the applied bias voltage $V_b$ and by subtracting the purely equilibrium term $\hat{\rho}_{\rm eq} = -\frac{1}{\pi} \int dE \, {\rm Im}\, \hat{G}^r_0(E) f(E)$:
\begin{eqnarray}\label{eq:rhoneq}
\lefteqn{Q_{\rm neq}  =  -\sum_{p} V_p {\rm Tr} \left[ \frac{\partial \hat{G}^r_0}{\partial q} \hat{\Gamma}_{p} \hat{G}^a_0 \hat{\Gamma} - i\frac{\partial \hat{G}^r_0}{\partial q} \hat{\Gamma}_{\alpha} \right]} \nonumber \\
&& -\sum_{p} V_p {\rm Im} \left\{ \int\limits_{-\infty}^{E_F} \!\! dE\, {\rm Tr} \, \left[\frac{\partial \hat{G}^{r}_0}{\partial q} \frac{\partial \hat{H}}{\partial V_{p}}-\frac{\partial \hat{G}^{r}_0}{\partial q}\frac{\partial \hat{\Sigma}^r_p}{\partial E} \right] \right\}.
\end{eqnarray}
Here $G^r_0(E)$ is the retarded GF at zero bias voltage and we assume zero temperature. The second sum in Eq.~\eqref{eq:rhoneq} is non-zero only for ${\bf T}_\perp \propto V_b$ where the integration over the Fermi sea is necessary to ensure the {\em gauge invariance} (i.e., invariance under a global potential shift $V_p \rightarrow V_p + U$) of ${\bf T}_\perp$ plotted in Fig.~\ref{fig:fig2}. Note that ${\bf T}_\perp  \propto V_b$
component is identically zero~\cite{Wang2011,Xiao2008a,Tang2010} in symmetric F$^\prime|$I$|$F MTJs, as confirmed by Fig.~\ref{fig:fig2}(b) using our general Eq.~\eqref{eq:rhoneq} rather than the usual~\cite{Xiao2008a} special choice $V_L=-V_R=-V_b/2$ applicable only to MTJs with identical F$^\prime$ and F layers.

We conclude by noting that one of the key experimental issues for STT in conventional MTJs is its control via finite bias voltage~\cite{Wang2011}. While our Eq.~(\ref{eq:central}) intrinsically takes into account finite bias voltage, the effective Hamiltonian Eq.~(\ref{eq:ti}) is too crude to describe the band structure of a real 3D TI material necessary for such calculations. Similarly, computation of ${\bf T}_\perp \propto V_b$ for N$|$TI$|$F semi-MTJ requires integration in the second term in Eq.~\eqref{eq:rhoneq} over the whole energy band so that our result for ${\bf T}_\perp$ is also crude. Reliable integration over energy or finite bias calculations necessitate  coupling of Eq.~(\ref{eq:central}) to NEGF-DFT formalism~\cite{Jia2011} to capture band structure and interface reconstruction, as well as self-consistent charge and spin densities across the junction.

\begin{acknowledgments}
F. M. and B. K. N. were supported by DOE Grant No. DE-FG02-07ER46374 and N. N. was supported by
Grant-in-Aids for Scientific Research (21244053) from the Ministry of Education, Culture, Sports, Science and
Technology of Japan, Strategic International Cooperative Program (Joint Research Type) from Japan Science and
Technology Agency, and also by Funding Program for World-Leading Innovative R\&D on Science and Technology (FIRST Program).
\end{acknowledgments}





\end{document}